# Temperature-induced phase transitions in the quantum Hall magnet of bilayer graphene

M. Tanaka[1], K. Watanabe[2], T. Taniguchi[3], K. Nomura[4], S. Tarucha[5], M. Yamamoto[5]

[1]*Department of Applied Physics, University of Tokyo, Bunkyo-ku, Japan.*

[2]*Research Center for Functional Materials, National Institute for Materials Science, Tsukuba-shi, Japan.*

[3]*International Center for Material Nanoarchitectonics, National Instutute for Material Science, Tsukuba-shi, Japan*

[4]*Department of Physics, Tohoku University, Sendai-shi, Japan.*

[5]*Center for Emergent Matter Science, RIKEN, Wako-shi, Japan*

The quantum Hall system can be used to study many-body physics owing to its multiple internal electronic degrees of freedom and tunability. While quantum phase transitions have been studied intensively, research on the temperature-induced phase transitions of this system is limited. We measured the pure bulk conductivity of a quantum Hall antiferromagnetic state in bilayer graphene over a wide range of temperatures and revealed the two-step phase transition associated with the breaking of the long-range order, i.e., the Kosterlitz–Thouless transition, and short-range antiferromagnetic order. Our findings are fundamental to understanding electron correlation in quantum Hall systems.

The quantum Hall state is one of the most strongly electronically correlated states owing to its quenched kinetic energy. When multiple internal electronic degrees of freedom exist, an exchange interaction stabilizes a many-body-ordered ground state if a one-particle Landau level (LL) is partially filled [1,2].

As it is characterized by energy gap opening and ordering, the quantum Hall state should have analogies with other correlated insulators such as Mott insulators and two-dimensional Moire flat band systems. Because both the interaction energy and one-particle energy of the quantum Hall state can be controlled by parameters such as the carrier density (filling factor), out-of-plane and in-plane magnetic field, and out-of-plane electric field, it can be a tunable experimental platform for investigating general correlated effects and phase transitions.

Although, the quantum phase transitions in quantum Hall states have been extensively studied both experimentally and theoretically, few studies have been conducted on temperature-induced classical phase transitions. This is because, theoretically, finite temperature behavior is much more difficult to investigate than zero-temperature behavior. Furthermore, experimentally, the coexistence of the bulk and edge states makes the temperature dependence of observables more complex than in homogeneous systems.

The zero-energy LL of bilayer graphene is a promising platform for studying temperature-induced phase transitions. It exhibits various ordered states owing to the interplay of spin, layer, and orbital degrees of freedom, and controllability of the layer degree of freedom by an out-of-plane external electric field (displacement field *D*) [3-34].

At $\nu = 0$ (half filling of the zero-energy LL), the canted antiferromagnetic (CAF) state is thought to be stabilized by the short-range Coulomb interaction under a small $D$, whereas the layer polarized (LP) state is favored under a large $D$ [3-25]. The ferromagnetic state is favored for enhanced Zeeman energy by a tilted magnetic field [11-13,21-24]. In this study, we focused on the CAF state, where the spins tend to align ferromagnetically within each layer and antiferromagnetically between the layers [21-23] (Fig. 1(a)). The spins tend to lie in the plane with a small canting along the out-of-plane magnetic field to minimize both the antiferromagnetic exchange energy and Zeeman energy. Therefore, the CAF state is almost an easy-plane antiferromagnet with U(1) symmetry. It is also thought to be stabilized in a double-layer semiconductor quantum well and the $\nu = 0$ state of monolayer graphene without staggered potential, where the layer degree of freedom in bilayer graphene is replaced with the sublattice degree of freedom.

Importantly, the CAF state does not have a zero-gap edge state unless the edge is a zigzag edge, owing to valley scattering at the edge. This simplifies the analysis of the temperature dependence of its bulk conductivity. In addition, the energy gap of the CAF state in bilayer graphene is much larger than that in a double-layer semiconductor quantum well [2], owing to the smaller separation between layers, which increases the phase transition temperature.

The CAF state has also attracted considerable interest for its unique electronic transport properties. Long-range spin current transport arising from the easy-plane antiferromagnetic order [35-38], a new kind of charge-neutral current originating from the spin-dependent layer polarization [39,40], and Kosteritz–Thouless (KT)-like critical behavior of the conductance [41,42] have been observed. In addition, recent theories indicate the easy-plane antiferromagnetism in magic-angle-twisted bilayer graphene, which is similar to the CAF, as an origin of its superconductivity [43].

Previously, the temperature dependence of the conductivity of the CAF state was measured in limited temperature ranges [11,14,41]. However, few discussions have been made on temperature-induced phase transitions, as will be discussed later.

In this study, we employed Corbino samples, which eliminate any type of edge transport to certainly measure the bulk conductivity in the CAF state and to study its temperature-induced phase transition. The observed nonmonotonic temperature dependence of the bulk conductivity implies a two-step phase transition, which is explained well by the two energy scales of the CAF state: the short-range Coulomb interaction and long-range Coulomb interaction energies.

Our measurements employed four samples: Corbino 1, Corbino 2, two-terminal (Supplemental Material), and a Hall bar (Supplemental Material). All the samples were dual-gated bilayer graphene encapsulated by hexagonal boron nitride (h-BN) (Fig.1b) and fabricated by the dry transfer technique and side contact (details are provided in the Supplemental Material). For Corbino 1 and Corbino 2, the dimensions of the active region covered with the top gate are the same. While a p-doped Si substrate is used as a back gate for Corbino 1, a graphite back gate is used for Corbino 2. For Corbino 2, the non-active region, which is not covered with the top gate, is heavily doped by the Si back gate. Therefore, most of the measured resistance originates from the active region. For Corbino 1, the resistance is the series resistance of the active and non-active regions. Because the CAF state is established at Vtg = 0 in Corbino 1, the active and non-active regions homogeneously become the CAF state under these conditions. This ensures the validity of the temperature dependence measurement, as mentioned later.

Although the CAF state generally has no ballistic edge state owing to valley scattering at the edge [3-12,21-25], there is possibility of diffusive edge transport owing to the hopping transport across sparsely existing zigzag edge regions [44]. The Corbino samples, which do not experience edge transport, allow for the measurement of pure bulk conductivity. We observed qualitatively similar temperature dependence in all samples above 6 K. Saturation of conductivity was observed below 6 K in Hall bar sample, which can be originated from edge transport or bulk hopping transport owing to sample dependent amount of impurities (Supplemental Material).

In Fig. 1c, we show the carrier density $n$ and displacement field $D$ dependence of the conductivity of Corbino 2, which was obtained from its gate voltage dependence (see Supplemental Material) under a perpendicular magnetic field $B$ = 9 T at temperature $T$ = 2.3 K. Periodic conductivity dips due to the formation of LLs were assigned to filling factors of $\pm 8, \pm 4, \pm 3, \pm 2, \pm 1$, and 0, as indicated in the figure. Focusing on $\nu = 0$ ($n = 0$), we found that the conductance dip vanished around $|D|$ = 0.16 V/nm. The two (separated) insulating states that appeared at $|D|$ < 0.16 V/nm and $|D|$ > 0.16 V/nm were assigned to the CAF state and the LP state, respectively [11,21-25]. The same feature was observed in Corbino 1, although some diagonal lines parallel to the axis of the top gate voltage arose from the non-active region (Supplemental Material).

The phase transition from the CAF state to the LP state was more clearly observed in the $D$ and $B$ dependences at $n = 0$ (Fig. 1d). The displacement field $D^*$ at the boundary between the CAF and LP regions linearly increases as B increases, which is quantitatively consistent with the results of a previous study [11]. Here, we convert $D^*$ into the energy unit $\Delta_{D^*}$ using the linear relationship between the displacement field and the energy gap at a zero magnetic field:

$$\Delta_{D^*} \equiv \Delta(D^*) \cong 130 \times D^*(\text{V/nm}) \text{ (meV)} \quad (1).$$

The function $\Delta(D) = 130$ meV/D (V/nm) is the energy gap induced by applying the displacement field $D$ at a zero magnetic field [15].

The dependence of $\Delta_{D^*}$ on B is shown in Fig. 2c. The physical meaning of $\Delta_{D^*}$ is the difference in the interaction energy between the CAF and LP states, which is overcome by the polarization energy at $D = D^*$.

Having confirmed the known gate-dependence property of the $\nu = 0$ quantum Hall state, we studied the temperature dependence of the conductivity at the center of the CAF state ($n = 0$, $D = 0$). Owing to the gate leakage problem of Corbino 2 at high temperatures, a wide range of temperature dependences were measured for the Corbino 1, two-terminal, and Hall bar samples. This measurement for Corbino 1 was not affected by its non-active region because the center of the CAF state is at $V$top gate = 0 and $V$back gate = 0; therefore, the entire sample was in the CAF state. The temperature dependence of the conductivity for all the three samples exhibited a nonmonotonic behavior (Fig. 2a and Supplemental Material). Corbino 1 at $B$ = 8 T behaved as an insulator below $T$ = 20 K, a metal at higher temperatures, and an insulator above $T$ = 80 K (Fig. 2a). We defined these three temperature regions as Ⅰ, Ⅱ, and Ⅲ, respectively. We defined the boundary temperature between I and II (II and III) as $T_{C1}$ ($T_{C2}$), where conductivity takes a local maximum (minimum), and are shown in Fig. 2c.

Figure 2b represents the Arrhenius plot of $\sigma$. The activation energies of regions I and III were defined as $\Delta_{\text{I}}$ and $\Delta_{\text{III}}$, respectively, and are shown in Fig. 2c. $T_{C2}$ and $\Delta_{\text{III}}$ are comparable to $\Delta_{D^*}$. $T_{C1}$ is much smaller than $T_{C2}$ and is larger than $\Delta_{\text{I}}$. This nonmonotonic $T$-dependence has been reported in previous studies [11,14,41]. However, its origin has not yet been determined. In a previous study, it was pointed out that nonmonotonicity can originate from the coexistence of bulk and edge states [14]. However, our results revealed that the non-monotonicity of the CAF state is due to an intrinsic bulk property.

We now consider the origin of the nonmonotonic $T$-dependence and physical significance of the characteristic temperatures based on the mean-field theory of quantum

Hall ferromagnetism. Generally, the energy gap of a quantum Hall FM system consists of three terms [21]:

$$E = E_1 + E_\circ + E_*,$$

$$E_1 = \mu_B B_{total} + \Delta(D),$$

$$E_\circ \simeq \frac{e^2}{4\pi\epsilon l_B} \propto \sqrt{B_\perp},$$

$$E_* \simeq \int dr^2 \left[\phi^*(r)\frac{e^2}{4\pi\epsilon a}\phi(r)\right]^2$$

$$= \frac{1}{l_B^2}\int dr'^2 \left[\phi^*(r')\frac{e^2}{4\pi\epsilon a}\phi(r')\right]^2 \quad (r' = r/l_B)$$

$$\propto 1/l_B^2 \propto B_\perp,$$

where $B_{total}$ ($B_\perp$) is total (out-of-plane) magnetic field, $\epsilon$ is the in-plane dielectric constant, $l_B$ is the magnetic length, $a$ is the lattice constant, and $\phi(r)$ is the wave function of the zero-th landau level. $E_1$ contains the Zeeman energy and polarization energy. At $D = 0$, the polarization energy is zero; therefore $E_1 = \mu_B B \simeq 0.7B[T]$ K. $E_\circ$ represents Coulomb interaction in a longer scale than lattice constant that is symmetric in the spin and valley space. This is proportional to the square root of the perpendicular magnetic field. Based on the theoretical calculation, $E_\circ \simeq 10\sqrt{B[T]}$ K is estimated [21]. $E_*$ is the lattice-scale short-range Coulomb interaction, which is valley asymmetric and proportional to the perpendicular magnetic field. A calculation gives $E_* \simeq 10\text{-}20\, B[T]$ K [21].

As $E_1$ and $E_\circ$ should be common in the CAF state and the LP state for the same parameters, the energy difference between these states $\Delta_{D*}$ is determined by $E_*$. This is consistent with the observation that $\Delta_{D*}$ is proportional to the perpendicular magnetic field. In our measurement range of the magnetic field, $E_* > E_\circ \gg E_1$ at $D = 0$; therefore, the energy gap of the CAF state is dominated by $E_*$. This is consistent with the results of the previous bias voltage-dependent transport experiment, where a linear dependence of the energy gap on the perpendicular magnetic field was observed [14].

In Fig. 2c, we show the theoretically expected values of $E_*$ and $E_\circ$ in red and blue shades, respectively. We find that $E_\circ$ is comparable to $T_{C1}$ and $E_*$ is comparable to $T_{C2}$, $\Delta_{\text{III}}$, and $\Delta_{D*}$, and. This indicates that $T_{C1}$ corresponds to the breaking of the quasi-long-range order (QLRO) and the higher characteristic temperatures ($T_{C2}$, $\Delta_{\text{III}}$, and $\Delta_{D*}$) correspond to the breaking of the short-range order (Fig. 3a).

We then considered the origin of the nonmonotonic temperature dependence of the conductivity based on this correspondence between $T_{C1}$ ($T_{C2}$) and long (short)-range Coulomb interaction energy. In region III, the temperature dependence of the conductivity is well fitted to the Arrhenius formula (Fig. 2b), and its activation energy is close to that of $T_{C2}$. Therefore, the conduction mechanism should be thermal excitation across the energy gap of the CAF state, which is mainly determined by the energy scale needed to break the local antiferromagnetic order ($E_*$).

In region I, the temperature dependence is roughly fitted to the Arrhenius formula with an activation energy smaller than $T_{C1}$ although it slightly deviates and exhibits weaker temperature dependence below 5 K. Because the temperature of region I is significantly lower than the energy gap, the hopping of carriers excited from the impurity states should be dominant.

In region II, the temperature dependence became metallic. As $T_{C1}$ corresponds to long-range Coulomb interaction, breaking of the QLRO is expected above $T_{C1}$. Because the CAF state has in-plane rotational symmetry, this order breaking is represented by the Kosterlitz–Thouless (KT) transition. Magnetic disorders promote spin flips of the carriers and increase the number of possible scattering processes, resulting in an increase in the scattering rate as the temperature increases. This type of conductivity reduction is generally observed in the paramagnetic-(anti)ferromagnetic transition of most magnetic materials and the magnetization flip of Ising ferromagnets which is known as butterfly shaped magnetoresistance [50].

Finally, we employ nonlocal transport measurement to get further insight on this scenario. We observed a decrease in nonlocal transport at the temperature close to $T_{C1}$, that can be related to the breaking of QLRO. A large nonlocal resistance has been observed in the CAF state, and its origin is thought to be the intrinsic flavor-dependent Hall conductivity arising from the spin and valley order [39]. Here, we measured the *T*-dependence of the nonlocal resistance in the range of 1.5 K to 50 K. We previously found that nonlocal resistance has a cubic scaling relationship with the local resistance at low temperatures, according to the semiclassical transport model with constant spin-valley Hall conductivity [40]. In Fig. 3b, we show $R_{NL}/R_{L3}$ as a function of temperature. It is nearly constant at low temperatures, indicating a constant spin-valley Hall conductivity. At higher temperatures, it drops and exhibits a dip (a black arrow in Fig. 3b). This drop indicates a drop in the spin-valley Hall conductivity. At higher temperatures, it increases as the temperature increases, indicating another mechanism of nonlocal transport, such as the thermal effect [40,49]. We defined the dip temperature as $T_{CNL}$, and plotted it in Fig. 2c. $T_{CNL}$ are comparable with $T_{C1}$.

For the nonlocal resistance to appear, the sign of the spin-valley Hall conductivity should be uniform throughout the entire sample, which requires homogeneous spin and valley ordering throughout the entire sample. Therefore, we need a QLRO whose correlation length is longer than the sample dimension (Fig. 3a). The decrease in nonlocal resistance from the cubic of the local resistance can be related to the breaking of the QLRO.

In summary, we observed a nonmonotonic temperature dependence of the conductivity in the CAF state characterized by two different energy scales. Based on the mean-field theory of quantum Hall ferromagnetism, we attribute these to the KT transition and the breaking of the local antiferromagnetic order. This is the first observation of a two-step temperature-induced phase transition of a quantum Hall magnet, which is theoretically argued for by the $\nu = 0$ quantum Hall state of monolayer graphene [42]. In Mott insulators, similar two-step phase transition associated with the breaking of the long-range and short-range antiferromagnetic orders is commonly observed [51-53], indicating the similarity between quantum Hall systems and correlated crystals. Our study could inform further studies of temperature-induced phase transitions in quantum Hall magnetic systems as gate-controllable experimental platforms.

M.Y. and S.T. acknowledge support from KAKENHI (Grant No. 26220710, 17H01138). K.W. and T.T. acknowledge support from the Elemental Strategy Initiative conducted by the MEXT, Japan (Grant Number JPMXP0112101001) and JSPS KAKENHI (Grant Numbers JP19H05790 and JP20H00354). K.N. acknowledge support from JSPS KAKENHI (Grant number JP20H01830) and CREST (Grant number JPMJCR18T2).

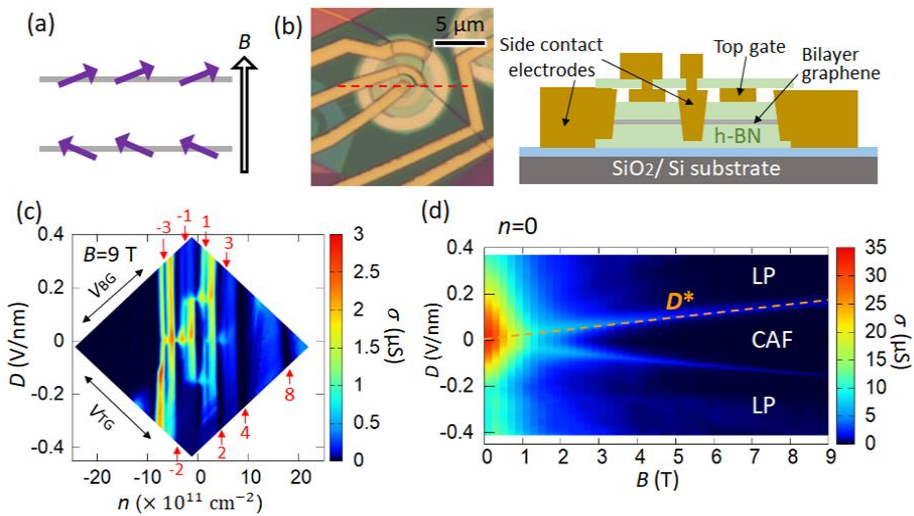

Fig. 1. CAF state, sample structure, and transport phase diagrams of bilayer graphene Landau levels. (a) Configuration of the spins in the CAF state in bilayer graphene. The gray lines are the top and bottom layers of the bilayer graphene. The spins are indicated by the purple arrows. (b) Optical microscope image of Corbino 1 (left) and a schematic cross-section along the broken red line (right). Bilayer graphene is encapsulated by high-quality hexagonal boron nitride (h-BN) crystals with a thickness of 30–50 nm and sandwiched between the gold top gate and p-doped Si back gate. (c) Plot of conductivity $\sigma$ versus carrier density $n$ and displacement field $D$ for Corbino 2 at $B$ = 9 T and $T$ = 2.3 K.

The double arrows indicate the axis of the top gate voltage $V_{TG}$ and back gate voltage $V_{BG}$ (see Supplemental Material for the conversion between $V_{TG}$–$V_{BG}$, and $n$–$D$). The red numbers are the filling factors assigned to the conductance dips in the $n$ axis. (d) Plot of $\sigma$ versus the perpendicular magnetic field $B$ and $D$ for Corbino 2 at $n = 0$ and $T = 2.3$ K. The orange broken line indicates the phase boundaries between the CAF and the LP regions.

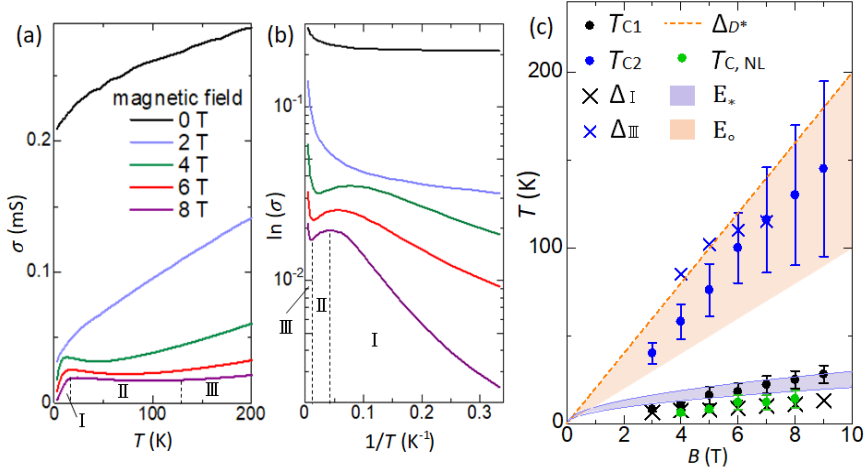

Fig. 2. Temperature dependence of the conductivity and parameters. (a, b) Standard plot (a) and Alenius plot (b) of the temperature dependence of the conductivity for Corbino 1 at $T = 2.3$–200 K for a magnetic field of 0, 2, 4, 6, and 8 T. The temperature regions separated by black broken lines are regions Ⅰ, Ⅱ, and Ⅲ for $B = 8$ T. (c) Magnetic field dependence of $T_{C1}$ (black dots), $T_{C2}$ (blue dots), $\Delta_Ⅰ$ (black cross), $\Delta_Ⅲ$ (blue cross), $\Delta_{D*}$ (orange broken line), $T_{CNL}$ (green dots), $E_*$ (blue shade), and $E_\circ$ (orange shade).

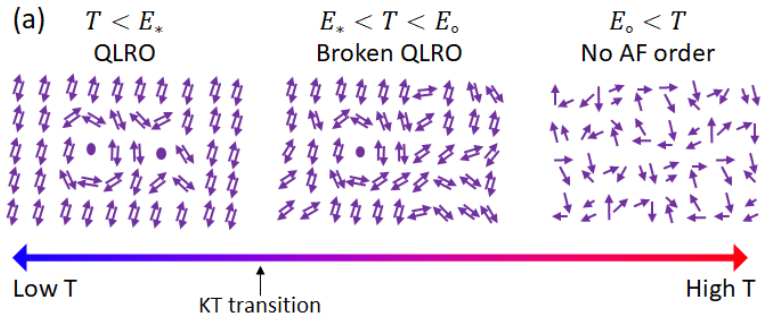

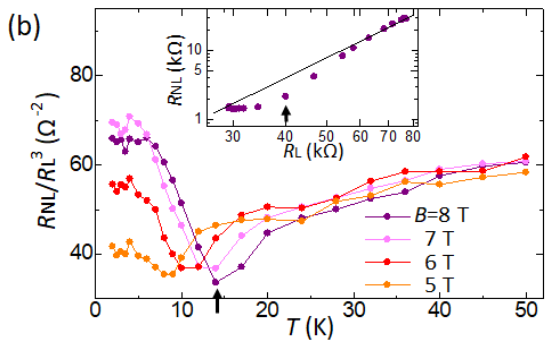

Fig. 4. Schematic illustrations of the ordering of the CAF state and nonlocal resistance. (a) Schematic of the phase transition in the CAF state. At low temperatures ($T < E_*$), vortices and antivortices are always bound, and the system has a QLRO. At $T = E_*$, QLRO is broken, and the correlation length begins to exponentially decrease as the temperature increases, but antiferromagnetic orders are still preserved locally. When the temperature is greater than the short-range Coulomb interaction $E_\circ$, the local antiferromagnetic order begins to vanish. (b) The plot of $R_{NL}/R_L^3$ as a function of temperature in the Hall bar sample. The arrows indicate $T_{CNL}$ for $B = 8$ T. The inset shows $R_{NL}$ as a function of $R_L$ at $B = 8$ T in the same temperature range. The black line and arrow indicate cubic dependence and $T_{CNL}$, respectively.